\documentclass{ifacconf}
\usepackage{graphicx}      % include this line if your document contains figures
\usepackage{natbib}        % required for bibliography
\usepackage{amsmath,amssymb,xcolor}%,fullpage}
\usepackage{tikz}
\usetikzlibrary{decorations.pathreplacing}
\usepackage{algorithm}
\usepackage[noend]{algpseudocode}
\newcommand{\bi}{\begin{itemize}}
\newcommand{\ei}{\end{itemize}}
\newcommand{\vo}[1]{\textcolor[HTML]{000000}{\boldsymbol{#1}}}
\newcommand{\x}{\vo{x}}
\newcommand{\f}{\vo{f}}
\newcommand{\g}{\vo{g}}
\newcommand{\h}{\vo{h}}
\newcommand{\vb}{\vo{v}}
\newcommand{\Vb}{\vo{V}}

\newcommand{\X}{\vo{X}}

\newcommand{\y}{\vo{y}}
\newcommand{\Mu}{\vo{\mu}}

\newcommand{\Q}{\vo{Q}}

\newcommand{\R}{\vo{R}}
\newcommand{\Y}{\vo{Y}}
\renewcommand{\H}{\vo{H}}
\newcommand{\I}{\vo{I}}
\newcommand{\D}{\vo{D}}

\newcommand{\A}{\vo{A}}
\newcommand{\B}{\vo{B}}

\newcommand{\w}{\vo{w}}
\newcommand{\W}{\vo{W}}
\newcommand{\F}{\vo{F}}

\newcommand{\real}{\mathbb{R}}
\newcommand{\Exp}[1]{\mathbb{E}\left[#1\right]}

\newcommand{\Sigp}{\vo{\Sigma}^{-}}

\newcommand{\Sigpp}{\vo{\Sigma}^{+}}

\newtheorem{theorem}{Theorem}
\newcommand{\eqnlabel}[1]{\label{eqn:#1}}
\newcommand{\figlabel}[1]{\label{fig:#1}}
\newcommand{\eqn}[1]{(\ref{eqn:#1})}
\newcommand{\fig}[1]{fig.(\ref{fig:#1})}
\newcommand{\Fig}[1]{Fig.(\ref{fig:#1})}
\usepackage{url}

\begin{document}
\begin{frontmatter}
\title{Optimal Sensing Precision in Ensemble and \\ Unscented Kalman Filtering\thanksref{footnoteinfo}}
\thanks[footnoteinfo]{This research was sponsored by Air Force Office of Scientific Research, Dynamic Data Driven Applications Systems grant FA9550-15-1-0071}
\author{Niladri Das \& Raktim Bhattacharya}
\address{Department of Aerospace Engineering,\\
        Texas A\&M University, College Station, Texas, USA.\\ (e-mail: niladridas,raktim@tamu.edu).}

\begin{abstract}
We consider the problem of selecting an optimal set of sensor precisions to estimate the states of a non-linear dynamical system using an Ensemble Kalman filter and an Unscented Kalman filter, which uses random and deterministic ensembles respectively. Specifically, the goal is to choose at run-time, a sparse set of sensor precisions for {active-sensing} that satisfies certain constraints on the estimated state covariance. In this paper, we show that this sensor precision selection problem is a semidefinite programming problem when we use $l_1$ norm over precision vector as the surrogate measure to induce sparsity. We formulate a sensor selection scheme over multiple time steps, for certain constraints on the terminal estimated state covariance.
\end{abstract}


\end{frontmatter}
%===============================================================================

\section{Introduction}
  In this paper, we focus on the problem of sensor design for non-linear stochastic discrete-time systems. Sensors are an integral part of a system, providing knowledge about system states, through state estimation (filtering), which can further be utilized to control the system. The problem of sensor design for a system, primarily addresses questions regarding, a) which type of sensor do we need, b) how accurate sensors do we need, and c) when and d) where do we use them, as mentioned in \cite{LI_2008}. 
The answers to the above problems explicitly depend upon, either the desired observability of the system, or the performance of the estimator and (or) the controller, or some performance metric of the system. This desired performance might also include minimizing energy consumption and total cost of operation or maximizing sensing accuracy or control performance, among various other metrics. In summary, sensor design strategy aims to strike a balance between the quality of sensing performance, sensing accuracy choice, and activation over space and (or) time.

  A considerable work on sensor design for state estimation focuses on addressing sensor selection problem, such as \cite{Joshi_2009,Zare_2018,1807.01739}, \textit{when sensor precisions are known}. A typical sensor selection problem either deals with choosing a minimal subset of sensors, from a set of available sensors that guarantees the state estimate covariance to be bounded, as in \cite{Tzoumas_2016}, or in \cite{Zhang_2017} where the authors minimizes the state estimate covariance when the cardinality of the sensor set is bounded. Owing to the combinatorial complexity of the problem, current methods are heuristic and developed in the Kalman filtering framework for linear Gaussian systems. There is also limited work on nonlinear state estimation which includes \cite{Chepuri_2014}, which focusses on choosing sparse sensor network with known sensor precisions.

  The focus of this paper \textit{is not on sensor selection, but on determining the sensor precisions}. Particularly, determining the least precise sensors for which state estimation is achieved with desired accuracy. This is achieved by solving an $l_1$ minimization problem. Once the sensor precisions are determined, existing sensor selection algorithms such as in \cite{Joshi_2009,Zare_2018,1807.01739}, can be applied to arrive at a reduced sensor set. However, due to the $l_1$ minimization, it is possible that the optimal solution assigns some of the precisions to zero, leading to sparsity in the sensor set. These sensors can be removed, indirectly addressing the sensor selection problem. Existing sensor selection algorithms can aid in further reducing the sensor set, possibly at the loss of estimation accuracy. In this work, we present sensor precision selection algorithm for nonlinear estimation based on ensemble Kalman filtering (EnKF) as discussed in \cite{Evensen_2003} and unscented Kalman filtering (UKF) in \cite{wan2000unscented}.

  Specifically, this paper addresses the problem of determining the accuracy (or precision) of a given dictionary of sensors, for a given upper-bound on the estimation accuracy. This problem has been addressed by \cite{LI_2008} for a linear continuous-time system, where the sensor precision and the control law were co-designed to achieve a specified closed-loop performance. The paper also presents a state-estimation problem, where sensor precisions were determined to achieve a certain estimation accuracy. In this paper, we look at a similar problem, but for a \textit{nonlinear discrete-time system}, with a user specified upper-bound on the estimation error covariance.

  This problem is important in many engineering applications where the choice and precision of sensors for state-estimation is not obvious. Examples of such applications include many large-scale spatio-temporal problems including space situational awareness where space objects are tracked using ground/space based sensor networks \cite{Das_2019}, structural health monitoring \cite{LI_2008}, environmental and climate monitoring \cite{Madankan_2014}, and distributed power-system monitoring \cite{Appasani_2017}, flow control applications \cite{da_Silva_2018}, and many other practical problems mentioned in \cite{Brockett, Zhang_2006, He_2006, Shi_2013, Han_2017, Chen_2017}.

  However, due to the constraints on the communication bandwidth and sensor battery life, it may not be desirable to have all the sensors report their measurements at all time instants or use the highest energy settings as in active sensing scenario such as in \cite{Chepuri_2015}. Therefore, determining what should be the least accuracy of each sensor to achieve a given accuracy in the state estimate becomes important from a practical point-of-view. Since precision of a sensor is explicitly related to its cost, solution to the sensor precision problem has economical implication.

  \noindent \textbf{Contributions of the Paper:}  In this paper we formulate a convex optimization problem to determine the optimal sensor precision for a given upper bound on the state estimation error covariance. This is presented for nonlinear discrete-time dynamical systems in EnKF and UKF frameworks. To the best of our knowledge, the sensor precision-selection problem for EnKF and UKF has not been addressed before. In this paper, the $l_1$ norm of the sensor precision is minimized, subjected to a convex constraint that guarantees the desired estimation error. This allows us to start with an over parameterization of the problem and determine a sparse solution via $l_1$ regularization. Therefore, the system designer can specify a dictionary of sensors with unknown sensor precisions and use the algorithm presented here to determine the optimal precision (and possibly eliminate a few sensors) to achieve the require estimation accuracy.

  It is important to understand that the true error covariance calculated using exact Bayesian update, might be different from that predicted by EnKF and UKF. When EnKF and UKF are used to approximate the exact Bayesian update, the approximate error covariance is guaranteed to be bounded, if the sensor precision selection algorithms presented in this paper are applied.

  \noindent \textbf{Notations:}  For a square matrix $\vo{M}$, let $\vo{M}^T$ denote its transpose. The variable $k\in \mathbb{Z}$ where $\mathbb{Z}$ is the set of integers, is used to index discrete time points; when used as subscript it refers to quantities taken at time $k$. The quantities ${\X}_{k}^{i-}$ and ${\X}_{k}^{i+}$ denotes  prior and posterior random variable associated with state $\X_{k}^i$, where the superscript $i$ denotes sample index. Observed value of the random variable $\Y_k$ is denoted by $\Y_k^o$. A positive definite matrix $\vo{M}$ is denoted by $\vo{M}\succ 0$. An identity  and a zero matrix of dimension $n\times n$ is denoted by $\vo{I}_{n\times n}$ and $\vo{0}_{n\times n}$ respectively. Random variable $\x$ which has a Gaussian distribution with mean $\vo{\mu}$ and covariance $\vo{\Sigma}$, is represented as $\x\sim \mathcal{N}(\vo{\mu},\vo{\Sigma})$. We represent the set of time indexed variable $\x_k$ as $\{\x_k\}$.

  \noindent \textbf{Layout of the Paper:}
The remainder of the paper is organized as follows. In Section 2, we present the system model along with its corresponding augmented model. In Section 3, we describe the EnKF and UKF filter models, leading to the problem formulation in Section 4, where we present the algorithms to solve the sensor precision selection problem. In Section 5, the proposed framework is applied to the Lorenz 1996 model. The paper finally concludes with Section 6.

\section{System Model}
  Consider an input/output discrete-time stochastic system modeled by,
\begin{align}
  \x_{k+1} &= \f_k(\x_k,\w_k),\eqnlabel{dynmodelm}\\
  \y_k &= \h_k(\x_k)+\vb_k\eqnlabel{measmodelm},
\end{align}
where $\f_k:\real^n\times \real^{n_w}\rightarrow \real^n$ represents the dynamics, $\h_k:\real^n\rightarrow \real^{n_y}$ is a measurement function, $\x_k\in \real^{n}$ and $\y_k\in \real^{n_y}$ are the state vector and the observation vector respectively, whereas $\w_k\in\real^{n_w}$ and $\vb_k\in\real^{n_y}$ are the process noise and measurement noise respectively. We assume that both $\{\w_k\}$ and $\{\vb_k\}$ are zero-mean, Gaussian, independent white random processes [$\w_k \sim \mathcal{N}(0,\Q_k), \vb_k \sim \mathcal{N}(0,\R_k),\Exp{\w_k\w_l^T}=\Q_k\delta_{kl}$, and $\Exp{\vb_k\vb_l^T}=\R_k\delta_{kl}$]. For sake of simplicity, the initial random variable $\x_0\sim \mathcal{N}(\vo{\mu}_0,\vo{\Sigma}_0)$ is independent of $\{\w_k\}$ and $\{\vb_k\}$. We assume that $\R_k$ is a diagonal matrix, representing the measurement noise covariance. The inverse of $\R_k$ is the referred to as the \textit{precision matrix}.

  In EnKF and UKF, the measurement data ($\y_k^o$) is used to determine the estimate of the state $\x_k$, which  minimizes the estimation variance.  We next introduce an \textit{augmented model}, based on \eqn{dynmodelm} and \eqn{measmodelm}, which aids in formulating a multi-step precision selection problem that satisfies the specified  performance criteria.

%\subsection*{Augmented Model}
  \noindent \textbf{Augmented Model:}  We consider each of the $q$ time steps $\{kq-q+1,...,kq\}$ of the system defined in \eqn{dynmodelm} \& \eqn{measmodelm} for $k\in \mathbb{Z}$, as a single time step

\begin{tikzpicture}
% Die Grundlinie:
\draw(0,0)--(1.5,0);
\node at (1.6,0) [circle,fill,inner sep=.5pt]{};
\node at (1.7,0) [circle,fill,inner sep=.5pt]{};
\node at (1.8,0) [circle,fill,inner sep=.5pt]{};
\draw(1.9,0)--(4,0);
\foreach \x/\xtext in {0/$1$,0.5/$2$,1/$3$,1.5/$4$,2/$q$}
    \draw(\x,1pt)--(\x,-1pt) node[below] {\xtext};
    \foreach \x/\xtext in {2.5/$q+1$,3/$$,3.5/$$,4.5/$2q$}
        \draw(\x,1pt)--(\x,-1pt) node[above] {\xtext};
\draw[decorate, decoration={brace}, yshift=2ex]  (0,0) -- node[above=0.4ex] {$q$ steps}  (1.9,0);
\node at (4.1,0) [circle,fill,inner sep=.5pt]{};
\node at (4.2,0) [circle,fill,inner sep=.5pt]{};
\node at (4.3,0) [circle,fill,inner sep=.5pt]{};
\draw(4.4,0)--(7.2,0);
\draw[decorate, decoration={brace, mirror}, yshift=-2ex]  (2.5,0) -- node[below=0.4ex] { $q$ steps}  (4.5,0);
\end{tikzpicture}

for the following augmented model:
\begin{align}
  \X_{k+1} = \F_k(\X_k,\W_k), \ \Y_k = \H_k(\X_k)+\Vb_k,\eqnlabel{augmodel}
\end{align}
where,
\begin{align}
  \X_k &:= [\x_{kq-q+1}^T,...,\x_{kq}^T]^T,\eqnlabel{augx}\\
  \Y_k &:= [\y_{kq-q+1}^T,...,\y_{kq}^T]^T,\nonumber\\
  \W_k &:=[\w_{kq-q+1}^T,...,\w_{kq+q-1}^T]^T\sim \mathcal{N}(\vo{0},\vo{\mathcal{Q}}_k),\nonumber\\
  \Vb_k&:=[\vb_{kq-q+1}^T,...,\vb_{kq}^T]^T\sim \mathcal{N}(\vo{0},\vo{\mathcal{R}}_k),\nonumber\\
  \vo{\mathcal{Q}}_k&:= \text{diag}([\Q_{kq-q+1},...,\Q_{kq+q-1}]),\eqnlabel{augQ}\\
    \vo{\mathcal{R}}_k&:= \text{diag}([\R_{kq-q+1},...,\R_{kq}]),\nonumber
  \end{align}
   denotes stacked random variables. Function $\F_k(.)$ can be recursively generated using $\vo{f}_i(.)$s. {It should be noted that the augmented model represents \textit{a $q$-step $q$-shift process, rather than a $q$-step sliding-window process.}}\\

  \noindent \textbf{Remark 1:}
In the rest of the paper, we only use the augmented state model and consequently time step $k$ denotes the batch of $q$ time points $\{kq-q+1,...,kq\}$, unless otherwise specified.

\section{Filter Models: EnKF and UKF}
  The filtering process for the augmented model \eqn{augmodel} consists of two sequential steps: dynamics update and measurement update. In EnKF, random samples are generated using Monte Carlo techniques, whereas the state distribution in UKF is represented by a Gaussian random variable (GRV) and is specified using a minimal set of carefully chosen deterministic sample points along with their associated weights, as shown in \cite{wan2000unscented}. The sensor-selection problem for each these filtering frameworks are presented in the next section.

\subsection{Dynamic Update for EnKF Model}
   Let $\vo{\mathcal{X}}_k^{+}\in\real^{nq\times N}$ be the matrix with $N$ number of \textit{posterior} samples $\X_k^{i+}$ at time $k$, i.e.
 $$\vo{\mathcal{X}}_k^{+} = \begin{bmatrix}\X^{1+}_k & \X^{2+}_k & \cdots & \X^{N+}_k \end{bmatrix}.$$
%\end{align}
The posterior mean from the samples is approximated as,
$$\Mu_k^{+} := \Exp{\X_k^{+}} \approx \frac{1}{N}\sum_{i=1}^{N}\X_k^{i+} =  \frac{1}{N}\vo{\mathcal{X}}_k^{+}\vo{1}_N,
$$
where $\vo{1}_N\in\real^N$ is a column vector of $N$ ones.
We define,
$$\bar{\vo{\mathcal{X}}}_k^{+}:=\begin{bmatrix}\Mu_k^{+} & \cdots & \Mu_k^{+}\end{bmatrix} = \Mu_k^+\vo{1}^T = \frac{1}{N}\vo{\mathcal{X}}_k^+\vo{1}\vo{1}^T,$$
then variance from the samples $\vo{\Sigma}_{xx,k}^{+}$ is,
\begin{align}
& \Exp{(\X_k^{i+}-\Mu_k^{+})(\X_k^{i+}-\Mu_k^{+})^T} \approx  \vo{\mathcal{X}}_k^+\A\vo{\mathcal{X}}_{k}^{+T}\eqnlabel{sigxx}.
\end{align}
where $$\A:=\left[\frac{1}{N-1}\left(\I_N-\frac{\vo{1}\vo{1}^T}{N}\right)\left(\I_N-\frac{\vo{1}\vo{1}^T}{N}\right)\right]$$
The state of each ensemble member at the next time step is estimated using the dynamics model:
  \begin{align}
    {\X}_{k+1}^{i-} &= \F_k({\X}_{k}^{i+},\W_k^{i}),\eqnlabel{sampledyn}
  \end{align}
  If applied to a linear system, this ensemble approach reduces the cost
  associated with the time propagation of the covariance matrix from
  $\mathcal{O}(n^3q^3)$ (classical KF) to $\mathcal{O}(n^2q^2N)$ (EnKF).

%  Because typical ensemble
%  sizes are no larger than $10^2$, the overall cost is usually reduced by
%  several orders of magnitude \cite{de_Castro_da_Silva_2017}.
    \subsection{Measurement update for EnKF Model}
  The ensemble members are corrected to minimize the error with respect to the measurements in the presence of noise and model uncertainties. A measurement update formulation proposed by \cite{evensen1996assimilation} is:
  \begin{align}
    {\X}_{k+1}^{i+} =& {\X}_{k+1}^{i-} + \Sigp_{xy,k+1}(\Sigp_{yy,k+1}+\vo{\mathcal{R}}_k)^{-1}\nonumber\\&\times(\Y^o_{k+1}-\vo{H}_{k+1}({\X}_{k+1}^{i-})+\boldsymbol{\epsilon}_k^i),
  \end{align}
where $\boldsymbol{\epsilon}_k^i$ is sampled from $ \mathcal{N}(\vo{0},\vo{\mathcal{R}}_k)$.
We define $\Sigp_{xy,k+1}$ as:
\begin{align}
\Sigp_{xy,k+1}=&\frac{1}{N-1}(\vo{\mathcal{X}}_{k+1}^--\bar{\vo{\mathcal{X}}}_{k+1}^-)\times \nonumber \\&(\H_{k+1}(\vo{\mathcal{X}}_{k+1}^-)-\H_{k+1}(\bar{\vo{\mathcal{X}}}_{k+1}^-))^T\eqnlabel{xysig},
\end{align}
 and  $\Sigp_{yy,k+1}$ is defined as:
\begin{align}
\Sigp_{yy,k+1}=& \frac{1}{N-1}\{\H_{k+1}(\vo{\mathcal{X}}_{k+1}^-)-\H_{k+1}(\bar{\vo{\mathcal{X}}}_{k+1}^-))\times \nonumber\\&(\H_{k+1}(\vo{\mathcal{X}}_{k+1}^-)-\H_{k+1}(\bar{\vo{\mathcal{X}}}_{k+1}^-))^T\}\eqnlabel{yysig}.
\end{align}

  \noindent  \textbf{Remark 2:} Equation \eqn{xysig} and \eqn{yysig} allows for direct evaluation of the nonlinear measurement function $\H_k(\x)$ in calculating the Kalman gain, which is shown in \cite{Tang_2014} to hold for unbiased measurement forecasts $\{\H_k(\X_k^{i-})\}$, which we assume to be true in our work.

The covariance update equation of the augmented model is:
\begin{align}
  \Sigpp_{xx,k+1} &= \Sigp_{xx,k+1} - \Sigp_{xy,k+1}(\Sigp_{yy,k+1}+\vo{\mathcal{R}}_{k+1})^{-1}\nonumber\\&\times{{\Sigp}_{xy,k+1}^T},\eqnlabel{sigR}
\end{align}
where $\Sigp_{xx,k+1}=\vo{\mathcal{X}}_{k+1}^-\A\vo{\mathcal{X}}_{k+1}^{-T}$.

\subsection{Dynamic Update for UKF Model}

  The dynamic update step from $k\rightarrow k+1$ starts with generating deterministic points  called $\sigma$ points. To capture the mean $_{a}\vo{\mu}_k^{+}$ of the augmented state vector $_{a}\X_k^{+}:=\begin{bmatrix}\X_k^{+} \\ \W_k\end{bmatrix}$, where $_{a}\X_k^{+}\in\mathbb{R}^{n_a}$ and $n_a = nq + n_wq$, as well as the augmented error covariance $_{a}\vo{\Sigma}_{xx,k}^{+}=\begin{bmatrix}\vo{\Sigma}_{xx,k}^+ & 0 \\0 &\vo{\mathcal{Q}}_k \end{bmatrix}$ 
the sigma points are chosen as 
\begin{align}
_{a}\X^{0+}_k &=  {_{a}\vo{\mu}_k^+}, \nonumber\\
_{a}\X^{i+}_k &=  {_{a}\vo{\mu}_k^+} + \Big(\sqrt{(n_a+\rho) _{a}\vo{\Sigma}_{xx,k}^+}\Big)_i, i=1,...,n_a, \nonumber\\
_{a}\X^{i+}_k &= {_{a} \vo{\mu}_k^+} - \Big(\sqrt{(n_a+\rho) _{a}\vo{\Sigma}_{xx,k}^+}\Big)_{i-nq} ,i=n_a+1,...,2n_a,\nonumber
\end{align}
with associated weights as 
\begin{align}
\vo{\omega}_{0}^{(m)} &= \rho/(n_a+\rho),\nonumber\\
\vo{\omega}_{0}^{(c)} &= \rho/(n_a+\rho) + (1-\alpha^2+\beta),\nonumber\\
\vo{\omega}_{i}^{(m)} &=  1/\{2(n_a+\rho)\}.\nonumber
\end{align}
The weight vectors are:
\begin{align}
\vo{\mathcal{W}}^{m} &= [\vo{\omega}_{0}^{(m)} \vo{\omega}_{1}^{(m)} ... \ \vo{\omega}_{2n_a+1}^{(m)}  ]^T,\nonumber\\
\vo{\mathcal{W}}^{c} &= [\vo{\omega}_{0}^{(c)} \vo{\omega}_{1}^{(c)} ... \ \vo{\omega}_{2n_a+1}^{(c)}  ]^T,\nonumber
\end{align}
where $\rho=\alpha^2(n_a+\kappa)-n_a$ is the scaling parameter, $\alpha$ is set to $0.001$, $\kappa$ is set to 0, and $\beta$ is 2 in this work. The term $\Big(\sqrt{(n_a+\rho) _a\vo{\Sigma}_{xx,k}^+}\Big)_i$ represents $i$th row of the matrix square root. The propagated state of each ensemble member at time $k+1$ is generated exactly as EnKF by using ${\X}_{k+1}^{i-} = \F_k({\X}_{k}^{i+},\W_k^{i})$, where $_a\X_{k}^{i+}:=\begin{bmatrix}\X_{k}^{i+}\\ \W_k^{i}\end{bmatrix}$.

But unlike EnKF, the corresponding prior mean and covariance at time $k+1$ are:
\begin{align}
{\vo{\mu}}_{k+1}^{-} &= \vo{\mathcal{X}}_{k+1}^{-}\vo{\mathcal{W}}^{m}\nonumber \\
\vo{\Sigma}_{xx,k+1}^{-}&=\vo{\mathcal{X}}_{k+1}^{-}\B_k\vo{\mathcal{X}}_{k+1}^{-T}\nonumber
\end{align}
where $\B_k:=\vo{L}\vo{L}^T$, $\vo{L}:=\left(\text{diag}(\vo{\mathcal{W}}^{c})-\vo{\mathcal{W}}^{c}\vo{1}_{2nq+1}^T\right)$.

We define the following terms which we use in the following measurement update phase of UKF.
\begin{align}
\vo{\mathcal{Y}}_{k+1}^-=\vo{H}(\vo{\mathcal{X}}_{k+1}^-), & \
\bar{\vo{\mathcal{Y}}}_{k+1}^{-}= \vo{\mathcal{Y}}_{k+1}^{-}\vo{\mathcal{W}}^{m}\vo{1}_{2nq+1}^T,\nonumber\\\bar{\vo{\mathcal{X}}}_{k+1}^{-}&= \vo{\mathcal{X}}_{k+1}^{-}\vo{\mathcal{W}}^{m}\vo{1}_{2nq+1}^T,\nonumber
\end{align}
where  $\vo{\mathcal{Y}}_{k+1}^{-} = \begin{bmatrix}\Y^{1-}_{k+1} & \Y^{2-}_{k+1} & \cdots & \Y^{(2nq+1)-}_{k+1} \end{bmatrix}$ and $\vo{\mathcal{X}}_{k+1}^{-} = \begin{bmatrix}\X^{1-}_{k+1} & \X^{2-}_{k+1} & \cdots & \X^{(2nq+1)-}_{k+1} \end{bmatrix}$

\subsection{Measurement Update for UKF}
  We calculate $\vo{\Sigma}_{xy,k+1}^{-}$ and $\vo{\Sigma}_{yy,k+1}^{-}$ as:
\begin{align}\vo{\Sigma}_{xy,k+1}^{-}&=(\vo{\mathcal{X}}_{k+1}^--\bar{\vo{\mathcal{X}}}_{k+1}^-)\times \text{diag}(\vo{\mathcal{W}}^{c})\nonumber \\ &\times(\vo{\mathcal{Y}}_{k+1}^{-}-\bar{\vo{\mathcal{Y}}}_{k+1}^{-})^T\eqnlabel{ukf1}\\
\vo{\Sigma}_{yy,k+1}^{-}&=(\vo{\mathcal{Y}}_{k+1}^{-}-\bar{\vo{\mathcal{Y}}}_{k+1}^{-})\times \text{diag}(\vo{\mathcal{W}}^{c})\nonumber \\ &\times(\vo{\mathcal{Y}}_{k+1}^{-}-\bar{\vo{\mathcal{Y}}}_{k+1}^{-})^T\eqnlabel{ukf2}
\end{align}
The covariance update equation is exactly same as \eqn{sigR}.

  \noindent \textbf{Remark 3:} Since the covariance update equation for EnKF and UKF are identical, this allows us to formulate a common precision selection algorithm that is presented next.

\section{Problem Formulation}
  We define precision matrix of measurement ($\vo{\mathcal{S}}_{k}$) as the inverse of the covariance matrix of the augmented measurement noise ($\vo{\mathcal{R}}_{k}$). We assume that the precision matrix $\vo{\mathcal{S}}_{k}$ is a diagonal matrix, with diagonal elements $\{\lambda_i\}$, where $i\in[1,...,qn_y]$. The sensor precision associated with the $i^{\text{th}}$ sensor is $\lambda_i$. Equation \eqn{sigR} for time step $k$, can be written as:
 \begin{align}
   &\Sigpp_{xx,k} = \Sigp_{xx,k} - \Sigp_{xy,k}(\Sigp_{yy,k}+\vo{\mathcal{S}}_{k}^{-1})^{-1}{{\Sigp}_{xy,k}^T}\nonumber\\
   &= \Sigp_{xx,k} - \Sigp_{xy,k}\{\Sigp_{yy,k}+\text{diag}([\lambda_1,...,\lambda_{qn_y}])^{-1}\}^{-1}\nonumber\\&\times{{\Sigp}_{xy,k}^T}
   \end{align}

   The $\lambda_i$'s are the control variables, that regulate the estimation error covariance matrix $\Sigpp_{xx,k}$, when we have the prior ensemble $\vo{\mathcal{X}}^{-}_k$ which is generated from $\vo{\mathcal{X}}^{+}_{k-1}$ using \eqn{sampledyn}. The augmented process noise $\vo{W}_k$ is generated by sampling from $\vo{\mathcal{Q}}_k$ in \eqn{augQ}.

Our objective is to design $\{\lambda_i\}$ such that $\vo{M}_q\Sigpp_{xx,k}\vo{M}_q^T$ is upper-bounded by a prescribed positive definite matrix $\boldsymbol{P}^d_{kq}$, where the matrix $\vo{M}_q:=[\vo{0}^1_{n\times n},\vo{0}^2_{n\times n},...,\vo{0}^{q-1}_{n\times n},\vo{I}_{n\times n}]$, is utilized to extract error covariance matrix of posterior estimate of $\x_{kq}$ from $\Sigpp_{xx,k}$. The matrix $\boldsymbol{P}^d_{kq}$ is the performance bound based on which we select sensor precisions.

\noindent\textbf{Remark 4:}  Although we use the augmented model in \eqn{augmodel}, the performance bound is on the covariance of the estimate of $\x_{kq}$.

\subsection{Optimal Sensor Precision}
  The solution to the sensor selection problem for EnKF and UKF is presented as the following theorem: \begin{theorem}
     \label{theorem1}
     The optimal precision of each of the sensors, $\boldsymbol{\lambda}_{k}:=[\lambda_1,...,\lambda_{qn_y}]$ at time $k$, which guarantees
   $  \vo{M}_q\Sigpp_{xx,k}\vo{M}_q^T \preceq \boldsymbol{P}^d_{kq}$, for given {prior ensemble $\vo{\mathcal{X}}^{+}_{k-1}$}, is obtained by solving the following semidefinite programming (SDP) problem,

   \begin{align}\boldsymbol{\lambda}_k^*= \min_{\boldsymbol{\lambda_k}:=[\lambda_1,...,\lambda_{qn_y}]^T}{||\boldsymbol{\lambda}_k}||_{1},\eqnlabel{l1min}\end{align}
   subject to,
   \begin{align}\begin{bmatrix}\boldsymbol{P}^d_{kq}+{\A} & {\B} \\ {\B}^T & {\D} \end{bmatrix}\succeq 0, \quad \lambda_i \geq 0, \ \forall i \in [1,...,qn_y],\eqnlabel{lmi}\end{align}
   where
   \begin{align}
   \A &:= -\vo{M}_q\Sigp_{xx,k}\vo{M}_q^T + \vo{M}_q{\Sigp}_{xy,k}\vo{\mathcal{S}}_{k}{{\Sigp}_{xy,k}^T}\vo{M}_q^T,\nonumber\\
    \B &:= \vo{M}_q\Sigp_{xy,k}\vo{\mathcal{S}}_{k},\nonumber \\
   \D &:= ({\Sigp}_{yy,k})^{-1}+\vo{\mathcal{S}}_{k},\nonumber \\
   \vo{\mathcal{S}}_{k}&:= \text{diag}([\lambda_1,...,\lambda_{qn_y}]). \nonumber
 \end{align}

The matrices $\Sigp_{xx,k},\Sigp_{xy,k},\Sigp_{yy,k}$ are calculated using the prior ensemble $\vo{\mathcal{X}}_k^{-}$ and the expected observations, $\H_k(\vo{\mathcal{X}}_k^{-})$ using \eqn{xysig} \& \eqn{yysig} for EnKF, or \eqn{ukf1} \& \eqn{ukf2} for UKF. We calculate $\vo{\mathcal{X}}_k^{-}$ from $\vo{\mathcal{X}}_{k-1}^{+}$ using \eqn{sampledyn} both for EnKF and UKF.
\end{theorem}

\begin{pf}
    \begin{align}
         \Sigpp_{xx,k} &= \Sigp_{xx,k} - \Sigp_{xy,k}(\Sigp_{yy,k}+\vo{\mathcal{R}}_k)^{-1}{{\Sigp}_{xy,k}^T}\nonumber\\
         &= \Sigp_{xx,k} - \Sigp_{xy,k}\{\vo{\mathcal{R}}_k^{-1}-\vo{\mathcal{R}}_k^{-1}\nonumber\\&\times[({\Sigp}_{yy,k})^{-1}+\vo{\mathcal{R}}_k^{-1}]^{-1}\vo{\mathcal{R}}_{k}^{-1}\}{{\Sigp}_{xy,k}^T}\nonumber\\
         &= \underbrace{\Sigp_{xx,k} - {\Sigp}_{xy,k}\vo{\mathcal{R}}_k^{-1}{{\Sigp}_{xy,k}^T}}_{-\vo{\hat{A}}}\nonumber\\&+\underbrace{\Sigp_{xy,k}\vo{\mathcal{R}}_k^{-1}}_{\vo{\hat{B}}}\underbrace{[({\Sigp}_{yy,k})^{-1}+\vo{\mathcal{R}}_k^{-1}]^{-1}}_{\vo{{D}}^{-1}}\nonumber \\ &\times\underbrace{\vo{\mathcal{R}}_k^{-1}{{\Sigp}_{xy,k}^T}}_{\vo{\hat{B}}^T} \nonumber\\
         \Sigpp_{xx,k} &= -{\vo{\hat{A}}} + {\vo{\hat{B}}}{\vo{{D}}}^{-1}{\vo{\hat{B}}}^{T},\nonumber
       \end{align}
       Collection the error covariance corresponding to posterior estimate of $\x_{kq}$.
       \begin{align}
              \vo{M}_q\Sigpp_{xx,k}\vo{M}_q^T &= -\vo{M}_q{\vo{\hat{A}}}\vo{M}_q^T + \vo{M}_q{\vo{\hat{B}}}{\vo{{D}}}^{-1}{\vo{\hat{B}}}^{T}\vo{M}_q^T \nonumber
       \end{align}
       Now,
       \begin{align}
         -{\A} + {\B}{\D}^{-1}{\B}^{T} \preceq \boldsymbol{P}^d_{kq} \nonumber \\
         \boldsymbol{P}^d_{kq}+{\A} - {\B}{\D}^{-1}{\B}^{T} \succeq 0,\eqnlabel{lmi123}
       \end{align}
       where $\vo{M}_q\vo{\hat{A}}\vo{M}_q^T:=\vo{{A}}$ and $\vo{M}_q\vo{\hat{B}}:=\vo{{B}}$.
       Since ${\D}\succ 0$ and $\boldsymbol{P}^d_{kq}+{\A}- {\B}{\D}^{-1}{\B}^{T} 	\succeq 0$, using Schur's complement we get the following,
   \begin{align}\begin{bmatrix}\boldsymbol{P}^d_{kq}+{\A} & {\B} \\ {\B}^T & {\D} \end{bmatrix}\succeq 0. \eqnlabel{theorem1proof}\end{align}
as the \textit{necessary and sufficient condition} for \eqn{lmi123} to be true.

   Matrices $\A,\B$, and $\D$ are linear in $\vo{\mathcal{R}}_k^{-1}$ or $\vo{\mathcal{S}}_{k}$. Equation \eqn{theorem1proof} is a linear matrix inequality (LMI) over $\vo{\mathcal{S}}_{k}$. The fact that the precision values are non-negative introduces the linear constraint $\lambda_i \geq 0$.
   The optimal precision is thus obtained by minimizing $\|\vo{\lambda}_k\|_1$, subject to the above LMI and linear constraint on $\vo{\lambda}_k$.

   \end{pf}

\begin{algorithm}
    \caption{Precision Selection }\label{euclid}
     \textbf{Input:} $\f_k(.),\g_k(.)$, $\h_k(.)$, $q$, $k$, $\vo{\mathcal{X}}_{k-1}^{+}$, $\vo{Q}_k$, $\boldsymbol{P}^d_{kq}$\\
    \textbf{Output:} A set $\vo\lambda\in \{\mathbb{R^+}^{(qn_y\times 1)}\cup \vo{0}_{qn_y\times 1}\}$ of sensor precisions.
    \begin{algorithmic}[1]
    \Procedure{}{}
    \State Calculate $\vo{\mathcal{Q}}_k$
    \State $\vo{\mathcal{X}}_{k-1}^{+}\rightarrow\vo{\mathcal{X}}_{k}^{-}$ using $\vo{F}_k(.)$, $\vo{W}_k\sim \mathcal{N}(\vo{0},\vo{\mathcal{Q}}_k)$
    \State Calculate $\Sigp_{xx,k},\Sigp_{xy,k},\Sigp_{yy,k}$
    \State Calculate $\vo{M}_q$
    \State Construct $\vo{\mathcal{S}}_{k}:= \text{diag}([\lambda_1,...,\lambda_{qn_y}])$
    \State Construct $\A,\B,\D$ matrices
    \State Solve SDP problem in \eqn{l1min}, \eqn{lmi}
    \EndProcedure
    \end{algorithmic}
\end{algorithm}

  \noindent \textbf{Remark 5}: Theorem \ref{theorem1} determines the optimal set of sensor precisions. The $l_1$ regularization induces sparseness in the solution. Therefore, if the optimization is performed on an over parameterized problem, i.e. with a large dictionary of sensors that considers all possible sensor choices, we expect the optimal solution $\vo{\lambda}^\ast$ to have entries that are zero. This indicates that those sensors are not needed to achieve the require state estimation accuracy and can be removed.

However, numerical solution to the sensor precision problem will result in small precision values that are not exactly zero. Those sensors can then be eliminated iteratively using theorem \ref{theorem1} with the reduced dictionary of sensors, discussed later in this paper. An upper bound on the precision for sensor(s) can also be incorporated in the optimization problem as a convex constraint over the argument space. For EnKF, covariance inflation \cite{Wu_2018} technique is used while calculating $\Sigp_{xx,k},\Sigp_{xy,k},\Sigp_{yy,k}$ matrices.

  \noindent \textbf{Remark 6:} Before solving the SDP problem, it is recommended to formulate the optimization problem with respect to normalized variables. This improves the numerical accuracy of the solution and make the optimal solution meaningful. For instance, in sensor precision selection for improving space-situational awareness, certain states are in the order of $10^3$ km and others are in radians. Therefore, normalization with respect to dynamics, error covariance, and sensor noise is required to avoid ill conditioning of covariance matrices and improve the efficacy of the proposed optimal sensor precision algorithm.

\subsection{Discussion: Sensor Pruning}
  As mentioned earlier, numerical solution of the $l_1$ regularization problem may assign small precision to certain $\{\lambda_{i}\}$s of $\vo{\lambda}$, which are not exactly zero and can possibly be removed without affecting the estimation quality. Therefore, a separate pruning process is required to reduce the set of sensors in the system. We define sensor pruning as choosing a subset of available sensors, which ensures that the covariance bound $\boldsymbol{P}^d_{kq}$ is satisfied. The rationale behind choosing a subset of the sensors is to eliminate the sensors whose precision requirement is too low compared to other sensors.

  An adhoc algorithm to address this has been presented in \cite{LI_2008}, where the calculated sensor precision vector $\vo{\lambda}^*$ is first sorted in ascending order. Iteratively, the smallest precision sensor are removed and the precisions are recalculated. This is continued till the problem becomes infeasible. The work of \cite{LI_2008} focusses on integrated design of controller and sensing architecture, without taking observability into consideration. However, in the context of state-estimation, observability condition must be addressed while sensor pruning. Other sensor selection algorithms proposed in \cite{Tzoumas_2016} and \cite{Zhang_2017} can also be investigated. However, the problem of sensor pruning becomes challenging for nonlinear system, and is the focus of our future work.

\section{Numerical Experiment}
In this section, we provide simulation results for the sensor precision selection algorithm for single time step update ($q=1$) and multiple time step update ($q=3$); also including the case where sensor precisions are constrained.

\subsection{Test Problem: The Lorenz (1996) model}
  The sensor precision selection scheme is applied to the Lorenz-96 (L96) model to test its validity, when an EnKF and an UKF filter are used for state estimation. The L96 model consists of $N_x$ equally spaced variables, $x_i$ for $i=1,...,N_x$, which are evolved in time using the set of differential equations:
\begin{align}
  \frac{dx_i}{dt} = (x_{i+1}-x_{i-2})x_{i-1}-x_i+F,\eqnlabel{L96}
  \end{align}
with cyclic boundaries: $x_{i+N}=x_i$ and $x_{i-N}=x_i$. The three terms in \eqn{L96} are analogous to advection, damping, and forcing. The system exhibits varying degrees of chaotic behaviors depending on the choices of $F$ and $N_x$. We fix $N_x$ and $F$ at 20 and 8 respectively, which leads to chaotic behavior in the system dynamics as shown in \cite{lorenz1996predictability}, and \cite{lorenz1998optimal}.

\subsection{Experimental Setup}
  In this experiment we consider no process noise, i.e. $\vo{\mathcal{Q}}_k=0$, but with initial condition uncertainty. Forward integration of \eqn{L96} is performed numerically using the fourth-order Runge-Kutta method with 20 internal stages for  $k\rightarrow k+1$, with a time step of 0.05 time units for each stage as shown in \cite{lorenz1996predictability}. We assume the following non-linear measurement model:
\begin{align}
  y_{i,k} = \frac{1}{1+e^{-x_{i,k}}} + v_{i,k} \eqnlabel{exmeas}
\end{align}
where $(.)_{i,k}$ denotes $i^{\text{th}}$ component of a vector at time point $k$, with measurement noise $\vb_k\sim \mathcal{N}(0,\R_k)$. The initial ensemble is generated from a multivariate Gaussian distribution with mean vector of size $qN_x$, whose elements are chosen randomly from $[0,F]$ and a random positive definite matrix ($\vo{\Sigma}_{{init}}$) of size $qN_{x}\times qN_{x}$ as the covariance, where $qN_{x}$ denotes the size of the augmented state vector. We use $2qN_x+1$ number of samples for both EnKF and UKF, for $q=1 \ \text{and} \ 3$.

To study the effects of state estimate covariance bounds on the optimal sensor precision values, we linearly vary the required error covariance bound from a factor of $0.9$ to $0.6$ of the initial covariance $\vo{\Sigma}_{{init}}$ as shown along the $x$-axis of the figures. For $q=1$ shown in \fig{fig1} and \fig{fig3}, 21 linearly varying bounds are considered within the interval of $[0.9,0.6]$, where as for $q=3$ shown in \fig{fig2}, \fig{fig4}, 7 linearly varying bounds are chosen from the same interval.  Measurement model in \eqn{exmeas} shows 20 different sensors, whose indices are plotted along the y-axis of each of the figures.

\subsection{Solving for Sensor Precisions}
  We use CVX software of \cite{cvx} to solve our SDP problem. CVX internally calls SeDuMi solver of \cite{sedumi}, a MATLAB implementation of the second-order interior-point methods. The $l_1$ norm minimization problem with LMI constraint yields desired precision values for the sensors shown in the figures as heat maps, with sensor precision ranges shown in the right y-axis. \Fig{fig1} and \fig{fig2} show optimal precisions required for EnKF for $q=1$ and $q=3$ respectively, satisfying prescribed covariance bounds. \Fig{fig3} and \fig{fig4}, are the corresponding plots for UKF. For $q=1$, sensor precisions are calculated to satisfy the covariance bound at the immediate next time instant. When $q=3$, sensor precision are calculated for consecutive 3 time instants to satisfy covariance bound on the state variable at the end of the time horizon. For EnKF, the sensor precisions are restricted to be below 15, whereas for UKF precisions are bounded above by 3, while solving the SDP problem. We see that the optimal solution results in high accuracy sensing only at the end of the time interval, with poor (or no) sensing within the interval.  However, this changes when upper limit on the precisions are reduced. In that case, we will see higher precision within the interval.

\begin{figure}[h]
\centering
\includegraphics[width=0.5\textwidth]{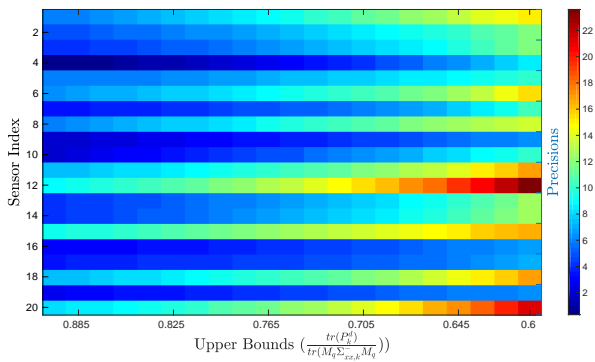}
\caption{Precision of sensors updated at each time step ($q=1$) for EnKF  without precision bounds}
\figlabel{fig1}
\end{figure}
\begin{figure}[h]
\centering
\includegraphics[width=0.5\textwidth]{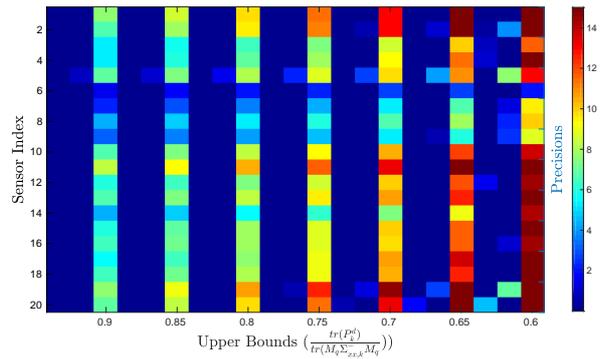}
\caption{Precision of sensors updated for 3 consecutive time step ($q=3$), with precision bounds for EnKF}
\figlabel{fig2}
\end{figure}
\begin{figure}[h]
\centering
\includegraphics[width=0.5\textwidth]{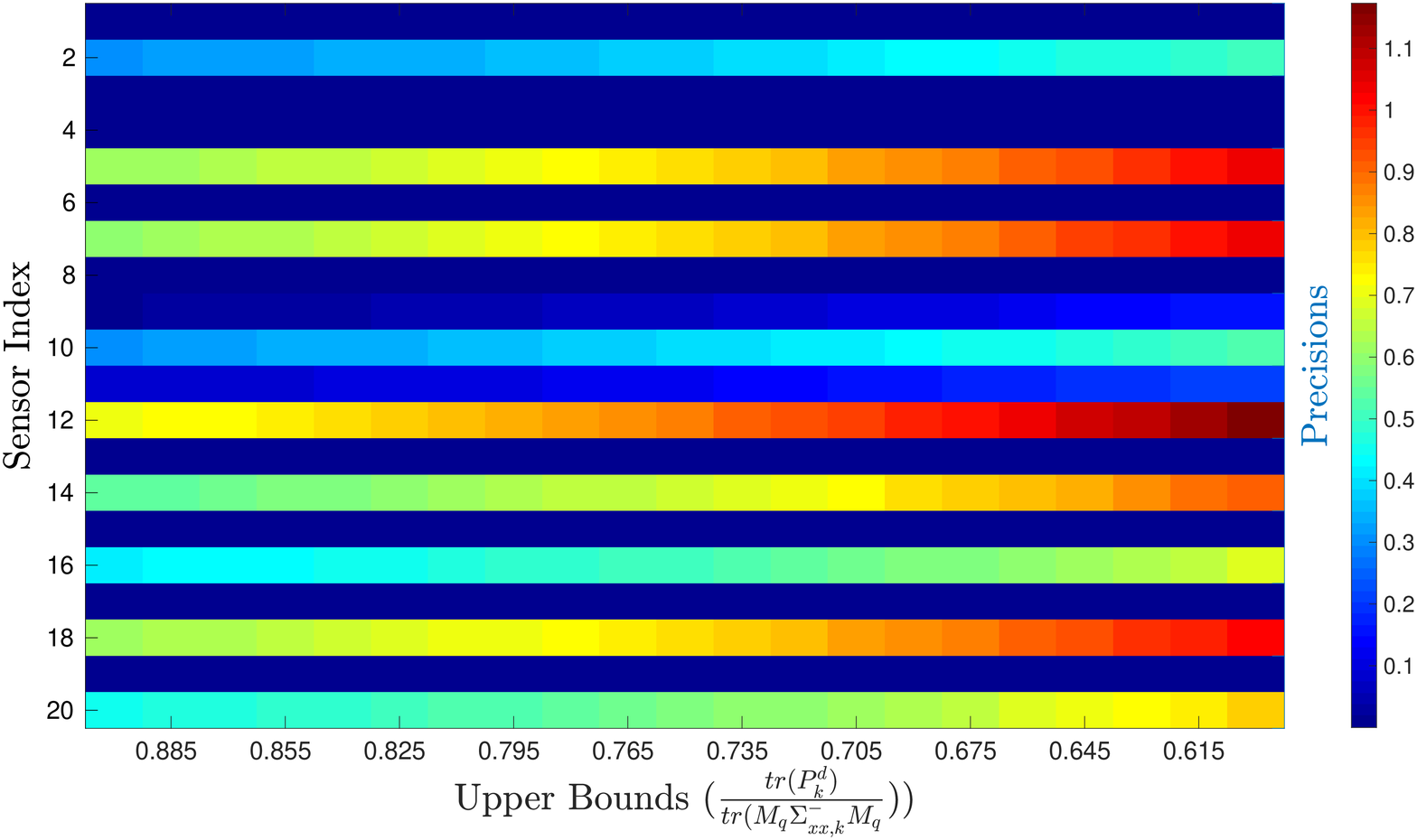}
\caption{Precision of sensors updated at each time step ($q=1$) for, for UKF without precision bounds }
\figlabel{fig3}
\end{figure}
\begin{figure}[h]
\centering
\includegraphics[width=0.5\textwidth]{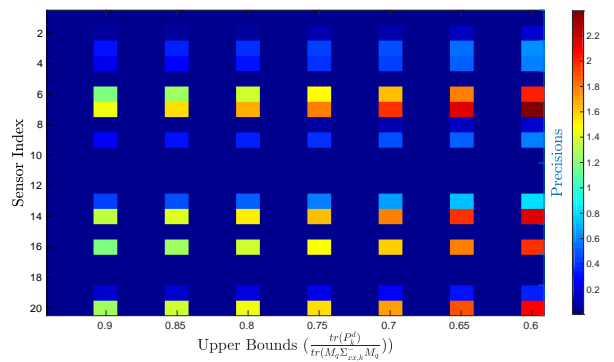}
\caption{Precision of sensors updated for 3 consecutive time step ($q=3$), for UKF with precision bounds }
\figlabel{fig4}
\end{figure}

Note that we get different values for the optimal precisions for EnKF and UKF. These also depend on the sample size and  covariance inflation parameter for EnKF, and choices of $\alpha, \beta, \kappa$ for UKF. It will be interesting to investigate the impact of these two frameworks on the sensor precision problem, and determine if one requires more precision than the other to arrive at the same estimation accuracy. These important questions will be addressed in our future work.

\section{Conclusion}
  In this paper, a new sensor precision selection problem for non-linear dynamical systems is presented in the framework of EnKF and UKF. The problem is shown to be convex, which can be easily solved using standard software such as CVX. The algorithm is applied to the Lorenz 1996 model of order 20 and results from both EnKF and UKF framework are presented. Sensor pruning, in the event of very small precisions in the optimal solution, is also discussed and methods to solve them are presented. Future work involves developing new sensor pruning algorithms for nonlinear systems, and also investigating impact of EnKF and UKF framework, along with other norm minimizations, on optimality and practicality.

\bibliography{IEEEexample}

\begin{thebibliography}{27}
\providecommand{\natexlab}[1]{#1}
\providecommand{\url}[1]{\texttt{#1}}
\providecommand{\urlprefix}{URL }
\expandafter\ifx\csname urlstyle\endcsname\relax
  \providecommand{\doi}[1]{doi:\discretionary{}{}{}#1}\else
  \providecommand{\doi}{doi:\discretionary{}{}{}\begingroup
  \urlstyle{rm}\Url}\fi

\bibitem[{Appasani and Mohanta(2017)}]{Appasani_2017}
Appasani, B. and Mohanta, D.K. (2017).
\newblock Optimal placement of synchrophasor sensors for risk hedging in a
  smart grid.
\newblock \emph{{IEEE} Sensors Journal}, 17(23), 7857--7865.
\newblock \doi{10.1109/jsen.2017.2742524}.
\newblock \urlprefix\url{https://doi.org/10.1109%2Fjsen.2017.2742524}.

\bibitem[{Brockett(1995)}]{Brockett}
Brockett, R. (1995).
\newblock Stabilization of motor networks.
\newblock In \emph{Proceedings of 1995 34th {IEEE} Conference on Decision and
  Control}. {IEEE}.
\newblock \doi{10.1109/cdc.1995.480312}.

\bibitem[{Chen et~al.(2017)Chen, Belabbas, and Ba{\c{s}}ar}]{Chen_2017}
Chen, X., Belabbas, M.A., and Ba{\c{s}}ar, T. (2017).
\newblock Optimal capacity allocation for sampled networked systems.
\newblock \emph{Automatica}, 85, 100--112.
\newblock \doi{10.1016/j.automatica.2017.07.039}.
\newblock \urlprefix\url{https://doi.org/10.1016%2Fj.automatica.2017.07.039}.

\bibitem[{Chepuri and Leus(2014)}]{Chepuri_2014}
Chepuri, S.P. and Leus, G. (2014).
\newblock Sparsity-promoting adaptive sensor selection for non-linear
  filtering.
\newblock In \emph{2014 {IEEE} International Conference on Acoustics, Speech
  and Signal Processing ({ICASSP})}. {IEEE}.
\newblock \doi{10.1109/icassp.2014.6854570}.
\newblock \urlprefix\url{https://doi.org/10.1109%2Ficassp.2014.6854570}.

\bibitem[{Chepuri and Leus(2015)}]{Chepuri_2015}
Chepuri, S.P. and Leus, G. (2015).
\newblock Sparsity-promoting sensor selection for non-linear measurement
  models.
\newblock \emph{{IEEE} Transactions on Signal Processing}, 63(3), 684--698.
\newblock \doi{10.1109/tsp.2014.2379662}.
\newblock \urlprefix\url{https://doi.org/10.1109%2Ftsp.2014.2379662}.

\bibitem[{da~Silva and Colonius(2018)}]{da_Silva_2018}
da~Silva, A.F.C. and Colonius, T. (2018).
\newblock Ensemble-based state estimator for aerodynamic flows.
\newblock \emph{{AIAA} Journal}, 56(7), 2568--2578.
\newblock \doi{10.2514/1.j056743}.
\newblock \urlprefix\url{https://doi.org/10.2514%2F1.j056743}.

\bibitem[{Das et~al.(2019)Das, Deshpande, and Bhattacharya}]{Das_2019}
Das, N., Deshpande, V., and Bhattacharya, R. (2019).
\newblock Optimal-transport-based tracking of space objects using range data
  from a single ranging station.
\newblock \emph{Journal of Guidance, Control, and Dynamics}, 1--13.
\newblock \doi{10.2514/1.g003796}.
\newblock \urlprefix\url{https://doi.org/10.2514%2F1.g003796}.

\bibitem[{Evensen(2003)}]{Evensen_2003}
Evensen, G. (2003).
\newblock The ensemble kalman filter: theoretical formulation and practical
  implementation.
\newblock \emph{Ocean Dynamics}, 53(4), 343--367.
\newblock \doi{10.1007/s10236-003-0036-9}.
\newblock \urlprefix\url{https://doi.org/10.1007%2Fs10236-003-0036-9}.

\bibitem[{Evensen and Van~Leeuwen(1996)}]{evensen1996assimilation}
Evensen, G. and Van~Leeuwen, P.J. (1996).
\newblock Assimilation of geosat altimeter data for the agulhas current using
  the ensemble kalman filter with a quasigeostrophic model.
\newblock \emph{Monthly Weather Review}, 124(1), 85--96.

\bibitem[{Grant and Boyd(2014)}]{cvx}
Grant, M. and Boyd, S. (2014).
\newblock {CVX}: Matlab software for disciplined convex programming, version
  2.1.
\newblock \url{http://cvxr.com/cvx}.

\bibitem[{Han et~al.(2017)Han, Wu, Zhang, and Shi}]{Han_2017}
Han, D., Wu, J., Zhang, H., and Shi, L. (2017).
\newblock Optimal sensor scheduling for multiple linear dynamical systems.
\newblock \emph{Automatica}, 75, 260--270.
\newblock \doi{10.1016/j.automatica.2016.09.015}.
\newblock \urlprefix\url{https://doi.org/10.1016%2Fj.automatica.2016.09.015}.

\bibitem[{He and Chong(2006)}]{He_2006}
He, Y. and Chong, E.K. (2006).
\newblock Sensor scheduling for target tracking: A monte carlo sampling
  approach.
\newblock \emph{Digital Signal Processing}, 16(5), 533--545.
\newblock \doi{10.1016/j.dsp.2005.02.005}.
\newblock \urlprefix\url{https://doi.org/10.1016%2Fj.dsp.2005.02.005}.

\bibitem[{Joshi and Boyd(2009)}]{Joshi_2009}
Joshi, S. and Boyd, S. (2009).
\newblock Sensor selection via convex optimization.
\newblock \emph{{IEEE} Transactions on Signal Processing}, 57(2), 451--462.
\newblock \doi{10.1109/tsp.2008.2007095}.
\newblock \urlprefix\url{https://doi.org/10.1109%2Ftsp.2008.2007095}.

\bibitem[{Li et~al.(2008)Li, Oliveira, and Skelton}]{LI_2008}
Li, F., Oliveira, M.C.D., and Skelton, R.E. (2008).
\newblock Integrating information architecture and control or estimation
  design.
\newblock \emph{{SICE} Journal of Control, Measurement, and System
  Integration}, 1(2), 120--128.
\newblock \doi{10.9746/jcmsi.1.120}.
\newblock \urlprefix\url{https://doi.org/10.9746%2Fjcmsi.1.120}.

\bibitem[{Lorenz(1996)}]{lorenz1996predictability}
Lorenz, E.N. (1996).
\newblock Predictability: A problem partly solved.
\newblock In \emph{Proc. Seminar on predictability}, volume~1.

\bibitem[{Lorenz and Emanuel(1998)}]{lorenz1998optimal}
Lorenz, E.N. and Emanuel, K.A. (1998).
\newblock Optimal sites for supplementary weather observations: Simulation with
  a small model.
\newblock \emph{Journal of the Atmospheric Sciences}, 55(3), 399--414.

\bibitem[{Madankan et~al.(2014)Madankan, Singla, and Singh}]{Madankan_2014}
Madankan, R., Singla, P., and Singh, T. (2014).
\newblock Optimal information collection for source parameter estimation of
  atmospheric release phenomenon.
\newblock In \emph{2014 American Control Conference}. {IEEE}.
\newblock \doi{10.1109/acc.2014.6858911}.
\newblock \urlprefix\url{https://doi.org/10.1109%2Facc.2014.6858911}.

\bibitem[{Shi and Chen(2013)}]{Shi_2013}
Shi, D. and Chen, T. (2013).
\newblock Optimal periodic scheduling of sensor networks: A branch and bound
  approach.
\newblock \emph{Systems {\&} Control Letters}, 62(9), 732--738.
\newblock \doi{10.1016/j.sysconle.2013.04.012}.
\newblock \urlprefix\url{https://doi.org/10.1016%2Fj.sysconle.2013.04.012}.

\bibitem[{Sturm(1999)}]{sedumi}
Sturm, J.F. (1999).
\newblock Using sedumi 1.02, a matlab toolbox for optimization over symmetric
  cones.
\newblock \emph{Optimization Methods and Software}, 11(1-4), 625--653.
\newblock \doi{10.1080/10556789908805766}.
\newblock \urlprefix\url{https://doi.org/10.1080/10556789908805766}.

\bibitem[{Tang et~al.(2014)Tang, Ambandan, and Chen}]{Tang_2014}
Tang, Y., Ambandan, J., and Chen, D. (2014).
\newblock Nonlinear measurement function in the ensemble kalman filter.
\newblock \emph{Advances in Atmospheric Sciences}, 31(3), 551--558.
\newblock \doi{10.1007/s00376-013-3117-9}.
\newblock \urlprefix\url{https://doi.org/10.1007%2Fs00376-013-3117-9}.

\bibitem[{Tzoumas et~al.(2016)Tzoumas, Jadbabaie, and Pappas}]{Tzoumas_2016}
Tzoumas, V., Jadbabaie, A., and Pappas, G.J. (2016).
\newblock Sensor placement for optimal kalman filtering: Fundamental limits,
  submodularity, and algorithms.
\newblock In \emph{2016 American Control Conference ({ACC})}. {IEEE}.
\newblock \doi{10.1109/acc.2016.7524914}.
\newblock \urlprefix\url{https://doi.org/10.1109%2Facc.2016.7524914}.

\bibitem[{Wan and Van Der~Merwe(2000)}]{wan2000unscented}
Wan, E.A. and Van Der~Merwe, R. (2000).
\newblock The unscented kalman filter for nonlinear estimation.
\newblock In \emph{Proceedings of the IEEE 2000 Adaptive Systems for Signal
  Processing, Communications, and Control Symposium (Cat. No. 00EX373)},
  153--158. Ieee.

\bibitem[{Wu and Zheng(2018)}]{Wu_2018}
Wu, G. and Zheng, X. (2018).
\newblock The error covariance matrix inflation in ensemble kalman filter.
\newblock In \emph{Kalman Filters - Theory for Advanced Applications}.
  {InTech}.
\newblock \doi{10.5772/intechopen.71960}.
\newblock \urlprefix\url{https://doi.org/10.5772%2Fintechopen.71960}.

\bibitem[{Zare and Jovanovic(2018)}]{Zare_2018}
Zare, A. and Jovanovic, M.R. (2018).
\newblock Optimal sensor selection via proximal optimization algorithms.
\newblock In \emph{2018 {IEEE} Conference on Decision and Control ({CDC})}.
  {IEEE}.
\newblock \doi{10.1109/cdc.2018.8619761}.
\newblock \urlprefix\url{https://doi.org/10.1109%2Fcdc.2018.8619761}.

\bibitem[{Zare et~al.(2018)Zare, Mohammadi, Dhingra, Jovanovi?, and
  Georgiou}]{1807.01739}
Zare, A., Mohammadi, H., Dhingra, N.K., Jovanovi?, M.R., and Georgiou, T.T.
  (2018).
\newblock Proximal algorithms for large-scale statistical modeling and optimal
  sensor/actuator selection.

\bibitem[{Zhang et~al.(2017)Zhang, Ayoub, and Sundaram}]{Zhang_2017}
Zhang, H., Ayoub, R., and Sundaram, S. (2017).
\newblock Sensor selection for kalman filtering of linear dynamical systems:
  Complexity, limitations and greedy algorithms.
\newblock \emph{Automatica}, 78, 202--210.
\newblock \doi{10.1016/j.automatica.2016.12.025}.
\newblock \urlprefix\url{https://doi.org/10.1016%2Fj.automatica.2016.12.025}.

\bibitem[{Zhang and Hristu-Varsakelis(2006)}]{Zhang_2006}
Zhang, L. and Hristu-Varsakelis, D. (2006).
\newblock Communication and control co-design for networked control systems.
\newblock \emph{Automatica}, 42(6), 953--958.
\newblock \doi{10.1016/j.automatica.2006.01.022}.
\newblock \urlprefix\url{https://doi.org/10.1016%2Fj.automatica.2006.01.022}.

\end{thebibliography}

\end{document}